\documentclass[10pt]{article}
\usepackage{amssymb}
\usepackage{latexsym}

\newcommand{\N}{{\mathbb N}}
\newcommand{\R}{{\mathbb R}}
\newcommand{\C}{{\mathbb C}}
\def\be{\begin{equation}}
\def\ee{\end{equation}}
\def\bea{\begin{eqnarray}}
\def\eea{\end{eqnarray}}
\def\Tr{{\rm \,Tr\,}}
\def\tr{{\rm \,tr\,}}
\def\d{{\,\rm d}}

\def\k{{\bf k}}

\def\i{{\bf i}}
\def\q{{\bf q}}
\def\n{{\bf n}}
\def\r{{\bf r}}
\def\p{{\bf p}}

\def\x{{\bf x}}
\def\y{{\bf y}}

\def\veps{\varepsilon}
\def\h2m{\frac{\hbar^2}{2m}}
\def\p0{{P_{\beta H^0_N}}}

\def\calH{{\cal H}}

\newtheorem{theorem}{Theorem}

\newtheorem{lemma}{Lemma}[section]
\newtheorem{prop}[lemma]{Proposition}
\newtheorem{coro}[lemma]{Corollary}

\begin{document}

\title{\large\bf Thermodynamic limit and proof of condensation for trapped bosons}
\author{Andr\'as S\"ut\H o\\
Research Institute for Solid State Physics and Optics\\ Hungarian Academy of
Sciences \\ P. O. Box 49, H-1525 Budapest\\ Hungary\\
E-mail: suto@szfki.hu}
\date{}
\maketitle
\thispagestyle{empty}
\begin{abstract}
\noindent
We study condensation of trapped bosons in the limit when the number of particles tends to infinity.
For the noninteracting gas we prove that there is no phase transition in any dimension, but in any
dimension,
at any temperature the system is 100\% condensated into the one-particle ground state.
In the case of an interacting gas we show that for a family of suitably scaled
pair interactions, the Gross-Pitaevskii scaling included,
a less-than-100\% condensation into a single-particle eigenstate, which may depend on the interaction
strength, persists at all temperatures.
\vspace{5mm}

\end{abstract}



\section{Introduction}

Bose-Einstein condensation (BEC) is one of the most fascinating collective phenomena occurring in Physics.
More than three quarters of a century after its discovery, the condensation of a homogeneous Bose gas
remains as enigmatic as ever, both experimentally and theoretically. Meanwhile, the experimental
realization of condensation in trapped atomic gases has opened new perspectives for the theory as well.
From the point of view of a mathematical treatment,
the trapped and the homogeneous systems are quite different, mainly due to
an energy gap above the -- at most finitely degenerate --
one-particle ground state of trapped Bose gases, implying that condensation
occurs into a localized state.
In the homogeneous gas the gap above the ground state vanishes in the thermodynamic limit. This makes
condensation a subtle mathematical problem already in the noninteracting system, and an
unsolved problem in the presence of any realistic interaction. The mathematical proof
of condensation in a trap shows no comparable subtlety, although
the gap endows the noninteracting gas with some peculiar properties, and condensation into a
localized state makes some sort of scaling of the interaction unavoidable.

A recent important development in the theory of trapped gases was obtained by Lieb and Seiringer \cite{LS}.
For a dilute interacting gas, in the limit when the particle number $N$ tends to infinity and the scattering
length $a$ to zero in such a way that $Na$ is fixed, these authors proved a 100\% BEC at zero temperature
into the Hartree one-particle wavefunction.

The aim of the present paper is to study BEC in deep traps, both in the free and in the interacting
cases. By a deep trap we mean a trap with an unbounded potential such that
the corresponding one-body Hamiltonian $H^0$ has a pure point spectrum and $\exp(-\beta H^0)$
is trace class for any positive $\beta$. Such a trap gives no possibility of escape to the particle
through thermal excitation. In Section 2 we prove a
condition on the potential so that it gives rise to a deep
trap.

In Section 3 we deel with the noninteracting gas
in the limit when the particle number, $N$, tends to infinity. We
show that asymptotically the total free energy is $N$ times the energy of the
one-particle ground state, plus an $O(1)$ analytic function of $\beta$. There is
no phase transition in any dimension $d\geq 1$, but the mean number
of particles in excited states remains finite as $N$ goes to infinity, whatever
be the temperature. So the density of the condensate is 1, condensation is 100\%
at all temperatures.

In Section 4 we use the results obtained for the noninteracting gas to prove the continuity of the phase
diagram as a function of the interaction strength.
In a first part, we define condensation into a one-particle state, and show that it is
equivalent to having the largest eigenvalue of the one-particle reduced density matrix
of order $N$. The second part of Section 4 contains the main result of the paper.
Here we prove a theorem on Bose-Einstein condensation in an interacting gas.
In particular, for a nonnegative interaction we obtain that, if the expectation
value of the $N$-particle interaction energy taken with the ground state of the noninteracting gas is
of the order of $N$, the occupation of at least
one of the low-lying eigenstates of the one-particle Hamiltonian, which may depend on the interaction
strength,
is macroscopic. This holds true in any dimension and at any finite temperature.
The result allows a finitely degenerate single-particle ground state (bosons with
spin) and is nonperturbative in the sense that it
does not depend on the size of the gap above the ground state.
The occupation of the subspace of one-particle ground states tends
to 100\% with the vanishing interaction strength.
In a corollary and in subsequent remarks
we describe a family of
nonnegative scaled interactions to which the theorem applies.
All these integrable pair interactions are weak, in the sense that their integral vanishes as
$1/N$ with an increasing number of particles.
Our examples include the Gross-Pitaevskii scaling limit in three dimensions and the opposite of
Gross-Pitaevskii scaling in one dimension.

\section{One-body Hamiltonian for deep traps}

The one-particle Hamiltonian we are going to use is
\be\label{H0}
H^0=-\h2m \Delta + V
\ee
on $L^2(\R^d)$, where the potential $V$ is chosen in such a way that $H^0$ has a pure point
spectrum with discrete eigenvalues of finite multiplicity and
$e^{-\beta H^0}$ is trace class, i.e. $\tr e^{-\beta H^0}<\infty$, for any $\beta>0$. This condition ensures the
finiteness of the one-particle free energy at any finite temperature $1/\beta$. We will refer to such a Hamiltonian
as a deep trap. For the sake of simplicity, we shall also suppose that the ground state of $H^0$
is nondegenerate, so that the eigenvalues of $H^0$ are
\be
\veps_0<\veps_1\leq\veps_2\leq\cdots\ .
\ee

A large family of potentials corresponding to deep traps is characterized by the following proposition.

\begin{prop}
Let $V:\R^d\to\R$ be bounded below and suppose that
\be\label{cond}
\lim_{r\to\infty}\frac{\ln(r/r_0)}{V(\r)}=0
\ee
for some $r_0>0$. Then $\tr e^{-\beta H^0}<\infty$ for all $\beta>0$.
\end{prop}

Condition (\ref{cond}) is sharp in the sense that, as the proof will show it, a 
central or cubic potential which increases logarithmically
leads to an exponentially increasing density of states and, therefore,
a diverging trace for small positive $\beta$.
Intuitively, the assertion of the proposition holds true because $\int\exp(-\beta V)\d\r<\infty$ for any $\beta>0$,
but the connection is not immediate.
We present two different proofs: The first uses the path integral representation of $\tr e^{-\beta H^0}$, while the
second is based on a semiclassical estimation of the eigenvalues.

\vspace{2mm}
\noindent
{\it First proof.} Given $\beta>0$, fix a $V_0>d/\beta$. Let $V_m=\inf V(\r)>-\infty$. If
(\ref{cond}) holds for an $r_0>0$ then it holds for any $r_0>0$. Choose $r_0$ so large that
\be\label{Vbound}
V(\r)\geq V_m+V_0\ln\frac{1}{2}\left(\frac{r}{r_0}+1\right)\quad\mbox{for all}\quad \r\in\R^d\ .
\ee
By the Feynman-Kac formula \cite{Gin},
\be\label{FK}
\tr e^{-\beta H^0}=\int\langle\r|e^{-\beta H^0}|\r\rangle \d\r
=\int P^\beta_{00}(\d\omega)\int e^{-\int_0^\beta V(\r+\omega(s))\d s} \d\r\ .
\ee
The first integral in the right member
goes over (continuous) paths $\omega$ in $\R^d$ such that $\omega(0)=\omega(\beta)=0$.
$P^\beta_{\x\y}(\d\omega)$ is the conditional Wiener measure, generated by $-\frac{\hbar^2}{2m}\Delta$,
for the time interval $[0,\beta]$,
defined on sets of paths with $\omega(0)=\x$ and $\omega(\beta)=\y$. In equation (\ref{FK}) we have made use of the
translation invariance of $P^\beta$. Let
\be
\|\omega\|_\beta=\sup_{0\leq s\leq \beta}|\omega(s)|\ .
\ee
The integral over $\r$ can be split in two parts. First,
\be
\int_{r<2\|\omega\|_\beta}e^{-\int_0^\beta V(\r+\omega(s))\d s} \d\r\
\leq e^{-\beta V_m}v_d(2\|\omega\|_\beta)^d
\ee
where $v_d$ is the volume of the $d$-dimensional unit ball. For $r>2\|\omega\|_\beta$, we use
(\ref{Vbound}) to obtain
\be
V(\r+\omega(s))\geq V_m+V_0\ln{r+2r_0\over 4r_0}\ .
\ee
After some algebra, this yields
\be
\int_{r>2\|\omega\|_\beta} e^{-\int_0^\beta V(\r+\omega(s))\d s} \d\r
\leq {e^{-\beta (V_m-V_0\ln 2)}s_d \over \beta V_0-d}(2r_0)^d\ .
\ee
Here $s_d$ is the surface area of the unit sphere in $\R^d$.
Putting the two parts together,
\be\label{upper}
\tr e^{-\beta H^0}\leq {e^{-\beta (V_m-V_0\ln 2)}s_d \over \beta V_0-d} \left({2r_0\over \lambda_\beta}\right)^d +
e^{-\beta V_m}v_d 2^d \int P^\beta_{00}(\d\omega)(\|\omega\|_\beta)^d\ ,
\ee
where we have substituted
\be
\int P^\beta_{00}(\d\omega)=\langle 0|e^{{\beta\hbar^2\over 2m}\Delta}|0\rangle=\lambda_\beta^{-d}\ ,
\ee
$\lambda_\beta=\hbar\sqrt{2\pi\beta/m}$ being the thermal de Broglie wave length. The second term on the right-hand side
of (\ref{upper}) is
finite: actually, every moment of the conditional Wiener measure is finite. Indeed, from the estimate
(see equations (1.14) and (1.31) of \cite{Gin})
\be
P^\beta_{00}(\|\omega\|_\beta>4\veps)\leq \frac{2^{2+d/2}}{\lambda_\beta^d}(m_d+n_d(\veps/\lambda_\beta)^{d-1})
e^{-\pi\veps^2/4\lambda_\beta^2}
\ee
where $m_d$ and $n_d$ depend only on the dimension $d$,
\be
\int P^\beta_{00}(\d\omega)(\|\omega\|_\beta)^k\leq
\frac{2^{2+d/2}}{\lambda_\beta^d}\sum_{n=0}^\infty (n+1)^k(m_d+n_d(n/4\lambda_\beta)^{d-1})
e^{-\pi n^2/64\lambda_\beta^2}<\infty\ .
\ee
This concludes the first proof.

\vspace{1mm}
\noindent
{\it Second proof.} We start, as before, by fixing $\beta>0$ and a $V_0>d/\beta$. For the sake of convenience,
now we choose $r_0$ so that
\be
V(\r)\geq V_m+V_0\ln\frac{1}{2}\left(\frac{r}{\sqrt{d}r_0}+1\right)\quad\mbox{for all}\quad \r\in\R^d\ .
\ee
The expression in the right member can still be bounded from below, due to the concavity of the square-root
and the logarithm. With the notation $\r=(x_1,\ldots,x_d)$, we find
\be
V(\r)\geq V_m-V_0\ln 2+\frac{V_0}{d}\sum_{i=1}^d \ln\left(\frac{|x_i|}{r_0}+1\right)\ .
\ee
Let
\be
h^0=-\frac{\hbar^2}{2m}\frac{\d^2}{\d x^2}+\frac{V_0}{d} \ln\left(\frac{|x|}{r_0}+1\right)\ .
\ee
Then
\be
H^0\geq V_m-V_0\ln 2+\sum_{i=1}^d h^0(i)\ ,
\ee
$h^0(i)$ acting on functions of $x_i$, and
\be
\tr e^{-\beta H^0}\leq e^{-\beta(V_m-V_0\ln 2)}\left(\tr e^{-\beta h^0}\right)^d\ .
\ee
Let $\lambda_n$, $n\geq 0$, be the eigenvalues of $h^0$ in increasing order. From Theorem 7.4 of \cite{Tit}, in the
case of a logarithmic potential, it follows that any $\lambda\in [\lambda_{n-1},\lambda_n]$ satisfies an equation
of the form
\be\label{lambn}
{\pi\hbar\over 2}(n+{1\over 2})=\int_0^X\sqrt{2m(\lambda-v(x))}\d x + O(\lambda)
\ee
where $X$ is defined by $v(X)=\lambda$. Dropping $O(\lambda)$, the solution is the $n\,$th semiclassical eigenvalue
according to the Bohr-Sommerfeld quantization.
For the true $n\,$th eigenvalue equation (\ref{lambn}) yields, after substituting
$v(x)=(V_0/d) \ln(|x|/r_0+1)$,
\be\label{lamn2}
\lambda_n=\frac{V_0}{d}\ln\left(n+\frac{1}{2}\right)+O(\ln\ln(n+3))\quad,\quad n=0,1,2,\ldots
\ee
So with a suitably chosen $c>0$ we obtain the bound
\be
\tr e^{-\beta h^0}=\sum_{n=0}^\infty e^{-\beta\lambda_n}
\leq\sum_{n=0}^\infty\frac{[\ln(n+3)]^{\beta c}}{(n+1/2)^{\beta V_0/d}}<\infty
\ee
which concludes the proof.

\vspace{2mm}
Observe that for $h^0$ and, thus, for the Hamiltonian $\sum_{i=1}^d h^0(i)$ the density of states can be
inferred from equation (\ref{lamn2}), and shows an exponential growth with the energy. This is origin of the
divergence of the trace for small $\beta$ in the case of logarithmically increasing potentials.

In the forthcoming proof of BEC at positive temperatures in interacting Bose gases, we shall make use 
of the following estimate.

\begin{prop}\label{2.2}
Let $e^{-\beta H^0}$ be trace class for every $\beta>0$, and suppose that $V$ is bounded below,
$\inf V=V_m$. Let $\varphi_j$ be the normalised eigenstate of $H^0$ belonging to the eigenvalue $\veps_j$.
Then
\be\label{supphi1}
e^{-\beta\veps_j}\|\varphi_j\|_\infty^2\leq e^{-\beta V_m}\left(\frac{m}{2\pi\hbar^2\beta}\right)^{d/2}
\ee
and
\be\label{supphi2}
\|\varphi_j\|_\infty^2\leq \left(\frac{em(\veps_j-V_m)}{d\pi\hbar^2}\right)^{d/2}\ .
\ee
\end{prop}

{\em Proof.}
\bea
e^{-\beta\veps_j}|\varphi_j(\r)|^2 &\leq& \sum_i e^{-\beta\veps_i}|\varphi_i(\r)|^2
=\langle\r|e^{-\beta H^0}|\r\rangle \nonumber\\
&=&\int P^\beta_{00}(\d\omega) e^{-\int_0^\beta V(\r+\omega(s))\d s}\leq e^{-\beta V_m}\lambda_\beta^{-d}
\eea
which, after taking the supremum on the left-hand side, is (\ref{supphi1}).
Multiplying by $e^{\beta\veps_j}$ and minimizing the right member with respect to
$\beta$ yields (\ref{supphi2}).

\vspace*{2mm}
We note that for sufficiently fast increasing potentials $H^0$ is ultracontractive and the
normalized eigenstates are uniformly bounded \cite{DS}, \cite{Ba}. In particular, this obviously holds true
for a particle confined in a rectangular domain with Dirichlet,
Neumann or periodic boundary conditions. All our results cover these cases.
However, in the present paper we need
only the weaker bound (\ref{supphi1}) on the eigenfunctions.

\section{Free Bose gas in a deep trap}\label{freeintrap}

$N$ noninteracting bosons in a deep trap are described by the Hamiltonian
\be
H^0_N=\sum_{i=1}^N H^0(i)=T_N+\sum_{i=1}^N V(\r_i)\qquad T_N=-{\hbar^2\over 2m}\sum_{i=1}^N \Delta_i\ .
\ee
We can consider $H^0_N$ directly in infinite space,
because $\exp(-\beta H^0_N)$ is a trace class operator on $L^2(\R^{dN})$.
Therefore, to perform a thermodynamic limit it remains sending $N$ to infinity.

Let $Z[\beta H^0_N]$ denote the canonical partition function for $N$ bosons.
We have the following.

\begin{prop}\label{freetrap1}
The limit
\be\label{freeinf}
\lim_{N\to\infty}e^{\beta N\veps_0}Z[\beta H^0_N]\equiv e^{-\beta F_0(\beta)}
\ee
exists, and $F_0(\beta)$ is an analytic function of $\beta$ for any $\beta>0$.
\end{prop}

\vspace{2mm}
\noindent
{\it Proof.} Let $n_j\geq 0$ denote the number of bosons in the $j\,$th eigenstate of $H^0$. Then
\be\label{Z0}
Z[\beta H^0_N]=\sum_{\{n_j\}:\sum n_j=N}e^{-\beta\sum n_j\veps_j}
=\sum_{N'=0}^Ne^{-\beta (N-N')\veps_0}\sum_{\{n_j\}_{j>0}:\sum n_j=N'}e^{-\beta\sum n_j\veps_j}\ .
\ee
Therefore
\be
e^{\beta N\veps_0}Z[\beta H^0_N]=\sum_{\{n_j\}_{j>0}:\sum n_j\leq N}e^{-\beta\sum n_j(\veps_j-\veps_0)}\ ,
\ee
so that
\bea
\lim_{N\to\infty}e^{\beta N\veps_0}Z[\beta H^0_N]
&=&\sum_{\{n_j\}_{j>0}:\sum n_j<\infty}e^{-\beta\sum n_j(\veps_j-\veps_0)}\nonumber\\
&=&\prod_{j=1}^\infty\sum_{n_j=0}^\infty e^{-\beta n_j(\veps_j-\veps_0)}
=\prod_{j=1}^\infty\frac{1}{1-e^{-\beta(\veps_j-\veps_0)}}
\eea
and
\be\label{F0}
\beta F_0(\beta)=\sum_{j=1}^\infty\ln(1-e^{-\beta(\veps_j-\veps_0)})\ .
\ee
To conclude, we need a lemma.

\begin{lemma}
Let $|a_n|<1$ and $\sum_{n=1}^\infty |a_n|<\infty$. Then $\sum_{n=1}^\infty\ln(1-a_n)$ is absolutely convergent.
\end{lemma}

\vspace{2mm}
\noindent
{\it Proof.} One can choose $N$ such that $|a_n|<1/2$ if $n\geq N$. Then
\bea
\sum_{n=N}^\infty|\ln(1-a_n)|&=&\sum_{n=N}^\infty|\sum_{l=1}^\infty \frac{a_n^l}{l}|\leq
\sum_{n=N}^\infty\sum_{l=1}^\infty \frac{|a_n|^l}{l}\nonumber\\
&=&\sum_{n=N}^\infty |a_n|\sum_{l=1}^\infty\frac{|a_n|^{l-1}}{l}\leq 2\ln 2
\sum_{n=N}^\infty |a_n|<\infty\nonumber
\eea
which proves the lemma.

\vspace{2mm}
Because $e^{-\beta H^0}$ is trace class for any
$\beta>0$, the conditions of the lemma hold for $a_n=\exp(-z(\veps_n-\veps_0))$ if $z\in\C$, Re$\,z>0$.
Thus, for any $\epsilon>0$,
$\sum_{n=1}^\infty\ln(1-\exp(-z(\veps_n-\veps_0)))$ is uniformly absolute convergent in the half-plane
Re$\,z\geq \epsilon$. Since every term is analytic,
the sum will be analytic as well. This finishes the proof of the proposition.

\vspace{2mm}
The peculiar feature of the infinite system is clearly shown by equation (\ref{freeinf}).
The total free energy of the gas is
\be
-\beta^{-1}\ln Z[\beta H^0_N]=N\veps_0+F_0(\beta)+ o(1)\ .
\ee
This means that there is no phase transition and
the free energy per particle of the infinite system
is $\veps_0$ at any temperature. Thus, at any $\beta>0$ the gas is in a low-temperature phase which is a nonextensive
perturbation of the ground state: {\it All but a vanishing fraction of the particles are in the condensate!} Below we
make this observation quantitative.

Let $\p0(A)$ denote the probability of an event $A$ according to the canonical Gibbs measure.
Let $N'=N-n_0$, the number of particles in the excited states of $H^0$.
First, notice that in the infinite system the probability that all the particles are in the ground state
is positive at any temperature: From equation (\ref{Z0}),
\be
\p0(N'=0)={e^{-\beta N\veps_0}\over Z[\beta H^0_N]}\rightarrow e^{\beta F_0(\beta)}\ \ {\rm as}\ \ N\to\infty
\ee
which tends continuously to zero only when $\beta\to 0$.
More precise informations can also be obtained.
For an integer $m$ between 0 and $N$, with Proposition \ref{freetrap1} we find
\be
\p0(N'\geq m)=
P_{\beta H^0_N}(N'=0)\sum_{\{n_j\}_{j>0}:m\leq\sum n_j\leq N}e^{-\beta\sum n_j(\veps_j-\veps_0)}\ .
\ee
A lower bound is obtained by keeping a single term, $n_1=m$, $n_j=0$ for $j>1$:
\be
\p0(N'\geq m)\geq P_{\beta H^0_N}(N'=0)e^{-\beta m(\veps_1-\veps_0)}\ .
\ee
If we replace $m$ by any increasing sequence $a_N$, this yields
\be\label{lower}
\liminf_{N\to\infty}\frac{1}{a_N}\ln \p0(N'\geq a_N)\geq -\beta(\veps_1-\veps_0)\ .
\ee
To obtain an upper bound, choose any $\mu$ with $0\leq\mu<\veps_1-\veps_0$. Then
\bea\label{uppm}
\p0(N'&\geq& m)\nonumber\\
&=&P_{\beta H^0_N}(N'=0)\sum_{N'=m}^N e^{-\beta\mu N'}
\sum_{\{n_j\}_{j>0}:\sum n_j=N'}e^{-\beta\sum n_j(\veps_j-\veps_0-\mu)}\nonumber\\
&\leq&P_{\beta H^0_N}(N'=0)\
e^{-\beta\mu m}\prod_{j=1}^\infty\frac{1}{1-e^{-\beta(\veps_j-\veps_0-\mu)}}\nonumber\\
&=&P_{\beta H^0_N}(N'=0)\ Q(\beta,\mu)\ e^{-\beta\mu m}
\eea
where $Q(\beta,\mu)$ is defined by the last equality. Notice that
$Q(\beta,0)=e^{-\beta F_0(\beta)}$.
The inequality has been obtained by first bounding $e^{-\beta\mu N'}$ above by $e^{-\beta\mu m}$ and then by
extending the summation over $N'$ from 0 to infinity.
Again, for $m=a_N\to\infty$,
\be\label{upp}
\limsup_{N\to\infty}\frac{1}{a_N}\ln \p0(N'\geq a_N)\leq -\beta\mu\ .
\ee
This being true for all $\mu<\veps_1-\veps_0$, it holds also for $\mu=\veps_1-\veps_0$, so the lower bound found in
(\ref{lower}) is an upper bound as well, and (\ref{lower}) and (\ref{upp}) together yield

\begin{prop}
If $0<a_N\leq N$ and  $a_N$ tends to infinity, then
\be
\lim_{N\to\infty}\frac{1}{a_N}\ln \p0(N'\geq a_N)= -\beta(\veps_1-\veps_0)\ .
\ee
\end{prop}

By the Borel-Cantelli lemma, inequality (\ref{uppm}) implies that
$N'$ is finite with probability 1 when $N$ is infinite. Moreover,
its expectation value is also finite: for any $\mu\in(0,\veps_1-\veps_0)$ we have
\be\label{avn'}
\langle N'\rangle_{\beta H^0_N}
\leq \frac{\p0(N'=0)\ Q(\beta,\mu)}
{\left(1-e^{-\beta\mu}\right)^2}
\ee
so that
\be\label{Nprime}
\lim_{N\to\infty}\langle N'\rangle_{\beta H^0_N}\leq\inf_{0<\mu<\veps_1-\veps_0}
\frac{Q(\beta,\mu)}{Q(\beta,0)}\frac{1}{(1-e^{-\beta\mu})^2}\ .
\ee
More generally, all moments of $N'$ remain finite as $N\to\infty$:
\be\label{nprimk}
\lim_{N\to\infty}\langle (N')^k\rangle_{\beta H^0_N}
\leq\inf_{0<\mu<\veps_1-\veps_0}
\frac{Q(\beta,\mu)}{Q(\beta,0)}\frac{\d^k}{\d(-\beta\mu)^k}
\frac{1}{1-e^{-\beta\mu}}\ .
\ee
This fact will be used in the proof of Theorem 2.

Let us summarize the results of this section:

\begin{theorem}
$N$ noninteracting bosons in a deep trap with eigen\-energies $\veps_0<\veps_1\leq\cdots$ have a free energy
$N\veps_0+F_0(\beta)+o(1)$, where $F_0$ is an analytic function of $\beta$ for any $\beta>0$. There is no phase
transition in any dimension, however, for any $d\geq 1$, the infinite
system is in a low-temperature phase ($T_c=\infty$):
At any finite temperature,
all but a finite expected number of bosons are in the one-particle ground state.
\end{theorem}

The conclusions about Bose-Einstein condensation are not modified if the ground state of $H^0$ is
degenerate. If
\be
\veps_0=\cdots=\veps_J<\veps_{J+1}\leq\cdots\ ,
\ee
we define $N'=N-(n_0+\cdots+n_J)$. Then, the earlier formulas remain valid if in the summations and
products $j>0$ is replaced by $j>J$, and $\veps_1-\veps_0$ is replaced by $\veps_{J+1}-\veps_0$. 
In particular, all moments of $N'$ are bounded and
we have a $100\%$ condensation into the finite dimensional subspace of ground states. 
This remark is relevant e.g. for bosons with an internal degree of freedom (spin).

\section{Condensation of interacting bosons}

\subsection{The order we are looking for}

Due to the pioneering work of Penrose \cite{Pen} and subsequent papers by
Penrose and Onsager \cite{PO} and Yang \cite{Yan}, it is generally understood and agreed
that Bose-Einstein condensation, from a mathematical point of view,
is an intrinsic property of the one-particle reduced
density matrix, $\sigma_1$, and  means that the largest eigenvalue
of $\sigma_1$, which is equal to its norm, $\|\sigma_1\|$,
is of the order of $N$. For the homogeneous gas the equivalence of this
physically not very appealing definition with the existence of an
off-diagonal long-range order, showing up
in the coordinate space representation (integral kernel) of $\sigma_1$, was
demonstrated in \cite{PO}. For a trapped gas it is intuitively more satisfactory
to define BEC as the accumulation of a macroscopic number of particles
in a single-particle state. The proof that this is meaningful, whether or not
there is interaction, and equivalent with $\|\sigma_1\|=O(N)$, is the subject
of this section.

Following the general setting  of \cite{Yan}, let $\sigma$ be a
density matrix, i.e., a positive operator of trace 1 acting in $\calH^N$, where $\calH$
is a one-particle separable Hilbert space. Permutations of the $N$ particles  
can be defined as unitary operators in $\calH^N$, and
$\sigma$ is supposed to commute with all of them.
The associated one-particle
reduced density matrix, $\sigma_1$, is a positive operator of trace $N$ in
$\calH$, obtained by taking the sum of the partial traces of $\sigma$ over the
$N\!-\!1$-\,particle subspaces: If
$\{\varphi_n\}_{n=0}^\infty$ is any orthonormal basis in $\calH$ and $\phi$ and
$\psi$ are any elements of $\calH$ then
\bea\label{sigma1}
(\phi,\sigma_1\psi) &\equiv& \sum_{j=1}^N\sum_{\{i_k\}_{k\neq j}}
(\varphi_{i_1}\!\cdots\!\varphi_{i_{j-1}}\phi\varphi_{i_{j+1}}\!\cdots\!
\varphi_{i_N},\sigma\,\varphi_{i_1}
\!\cdots\!\varphi_{i_{j-1}}\psi\varphi_{i_{j+1}}\!\cdots\!\varphi_{i_N})\nonumber\\
&=&N\sum_{i_2,\ldots,i_N}
(\phi\varphi_{i_2}\cdots\varphi_{i_N},\sigma\,
\psi\varphi_{i_2}\cdots\varphi_{i_N})
\eea
because of the permutation-invariance of $\sigma$. In (\ref{sigma1}) the
summation over each $i_k$ is unrestricted and the matrix elements of $\sigma$
are taken with simple (non-symmetrized) tensor products ($\otimes$ omitted).

Let $\varphi_0$ be any normalized element of $\calH$.
We define the {\em mean} (with respect to $\sigma$)
{\em number of particles occupying $\varphi_0$} as follows.
We complete $\varphi_0$ into an orthonormal basis $\{\varphi_n\}_{n=0}^\infty$
of $\calH$. In $\calH^N$ we use the product basis
\be\label{prodbase}
\{\Phi_\i=\varphi_{i_1}\otimes\cdots\otimes\varphi_{i_N}|\i=(i_1,\cdots,i_N)\in\N^N\}\ .
\ee
To $\varphi_0$ and $\Phi_\i$ we assign
\be\label{nvarphi}
n[\varphi_0](\i)\equiv\sum_{j=1}^N |(\varphi_0,\varphi_{i_j})|^2=\sum_j\delta_{i_j,0}\ ,
\ee
which is the number of particles in the
state $\varphi_0$ among $N$ particles occupying the states $\varphi_{i_1},\ldots,\varphi_{i_N}$,
respectively. 
We can use (\ref{nvarphi}) to define $n[\varphi_0]$ as a positive operator in $\calH^N$. Alternately,
we can interprete $(\Phi_\i,\sigma\Phi_\i)$ as the probability of $\Phi_\i$ and
$n[\varphi_0]$ as a random variable over this probability field. In any case, the
mean value of $n[\varphi_0]$ with respect to $\sigma$ is
\bea\label{nav}
\langle n[\varphi_0]\rangle_\sigma &\equiv& \Tr n[\varphi_0]\sigma
=
\sum_{i_1,\ldots,i_N}\sum_{j=1}^N\delta_{i_j,0}
(\Phi_\i,\sigma\Phi_\i)
\nonumber\\
&=&
\sum_{j=1}^N\sum_{i_j}\delta_{i_j,0}\sum_{\{i_k\}_{k\neq j}}
(\Phi_\i,\sigma\Phi_\i)
=
\frac{1}{N}\sum_{j=1}^N\sum_{i_j}\delta_{i_j,0}
(\varphi_{i_j},\sigma_1\varphi_{i_j})\nonumber\\
&=&
\sum_{i=0}^\infty\delta_{i,0}(\varphi_{i},\sigma_1\varphi_{i})
=(\varphi_0,\sigma_1\varphi_0)\ ,
\eea
an intrinsic quantity independent of the basis. Reading equation (\ref{nav}) in the
opposite sense, we find that, whether or not there is interaction,
{\em the physical meaning of $(\varphi_0,\sigma_1\varphi_0)$ is the average number of particles
in the single particle state $\varphi_0$.}
Since $\|\sigma_1\|=\sup_{\|\varphi\|=1}(\varphi,\sigma_1\varphi)$, we obtained
the following result.

\begin{prop}

There is BEC in the sense that $\lim_{N\to\infty}\|\sigma_1\|/N>0$ if and only if there exists
a macroscopically occupied $\varphi_0\in\calH$ (which may depend on $N$), i.e.
$\lim_{N\to\infty}\langle n[\varphi_0]\rangle_\sigma/N>0$.
\end{prop}

The proposition is valid with obvious modifications also in the homogeneous case.
The choice of the macroscopically occupied single particle state is
not unique.
Highest occupation is obtained for the dominant eigenvector, $\psi_{\sigma_1}$, of $\sigma_1$ (the one
belonging to the largest eigenvalue),
in which case $\langle n[\psi_{\sigma_1}]\rangle_\sigma=\|\sigma_1\|$.
Any other state having a nonvanishing overlap in the limit $N\to\infty$
with $\psi_{\sigma_1}$ can serve for proving BEC. We can even find an infinite
orthogonal family of vectors, all having a nonvanishing asymptotic overlap with $\psi_{\sigma_1}$.
One can speak about a generalized condensation \cite{BLP} only when the occupation of
more than one eigenstate of $\sigma_1$ becomes asymptotically divergent.
In the noninteracting
gas $\psi_{\sigma_1}$ is just the ground state of the one-body Hamiltonian.

The homogeneous gas represents a particular case. Namely,
$\psi_{\sigma_1}(\r)=\psi^L_{\k=0}(\r)\equiv 1/L^{d/2}$ for any translation invariant interaction,
if the gas is confined in a cube of side $L$ and
the boundary condition is periodic. Indeed, in this case
$\sigma_1$ is diagonal in
momentum representation, therefore $\psi^L_{\k}(\r)=e^{i\k\cdot\r}/L^{d/2}$ are its eigenstates. On the
other hand, the integral kernel $\langle\r|\sigma_1|\r'\rangle$ is positive
(now we speak about $\sigma\sim\exp(-\beta H)$ in the bosonic subspace or
$\sigma=|\Psi\rangle\langle\Psi|$ where $\Psi(\r_1,\ldots,\r_N)$ is a translation invariant positive
symmetric function), and by a suitable generalization of the Perron-Frobenius
theorem (e.g. \cite{KR}) the constant vector must be the dominant eigenvector. 
This is presumably the only case
when the ground state of the one-body Hamiltonian remains the dominant eigenvector
of the one-particle reduced density matrix for the interacting system, yet there exists
no proof of a macroscopic occupation of this state in the presence of interactions
(unless a gap is introduced in the excitation spectrum \cite{LVZ}).

In the case of
a trapped gas we do not generally know the dominant eigenvector
of $\sigma_1$. However, we can carry through
the proof by the use of the low energy eigenstates of $H^0$.

\subsection{Interacting bosons in a deep trap}

In this section we ask about condensation of interacting bosons in a deep trap.
Let $U_N:\R^{dN}\to\R$ be a symmetric function of $\r_1,\ldots,\r_N$ which is bounded below, and define
\be
H_N=H^0_N+U_N\ .
\ee
We can consider $H_N$ directly in infinite space,
because $\exp(-\beta H_N)$ is a trace class operator on $L^2(\R^{dN})$. So as
in Section \ref{freeintrap}, the thermodynamic limit
means $N$ tending to infinity.
The canonical partition function
will be denoted by $Z[\beta H_N]$.
The density matrix is
\be
\sigma=P^+_Ne^{-\beta H_N}/Z[\beta H_N]
\ee
where $P^+_N=(1/N!)\sum_{\pi\in S_N}\pi$ is the orthogonal projection to the symmetric subspace
of $\calH^N$
and
$Z[\beta H_N]=\Tr P^+_Ne^{-\beta H_N}$.

We want to prove the persistence of BEC in the presence
of interaction, that is, the continuity of the low-temperature phase as $U_N$ increases from zero to a
finite strength. This will be achieved
by proving macroscopic occupation of at least one low-lying eigenstate of
$H^0$, which may depend on the interaction strength.
We cannot expect, and will not obtain, a 100\% condensation
because the overlap of any eigenstate of $H^0$
with $\psi_{\sigma_1}$ must be smaller than 1.
(The 100\% condensation \cite{LS} into $\phi_{\rm GP}$, the minimizer of the Gross-Pitaevskii
functional, found for the ground state of interacting gases in the dilute limit
in locally bounded traps,
means that $(\phi_{\rm GP},\psi_{\sigma_1})\to 1$ as $N\to\infty$. In this case
$\sigma=|\Psi\rangle\langle\Psi|$, where $\Psi$ is the unknown ground state.)

In the proof of the next theorem we use the basis of the $H^0$ eigenstates, given by
$H^0\varphi_j=\veps_j\varphi_j$, and the
symmetric and normalized eigenstates of $H^0_N$: $|\n\rangle=|n_0,n_1,\ldots\rangle$ is the symmetrized
product state of norm 1 containing $n_j$ times the factor $\varphi_j$, where $\sum n_j=N$.
For the sake of clarity, we restrict the discussion to the case when the ground state of $H^0$ is
nondegenerate, and use the notation $\Phi_0$ for the (unique) ground state of $H^0_N$, given by $n_0=N$ and
$n_j=0$, $j>0$. Extension to the case of spin- or spatial degeneracy is straightforward.

\begin{theorem}\label{thminter}
Let
\be\label{cond1}
L(U)\equiv\limsup_{N\to\infty}\frac{1}{N}[(\Phi_0,U_N\Phi_0)-\inf U_N]\ .
\ee
For any $d\geq 1$ the following hold true.
\\
(i) If $L(U)<\infty$,
at zero temperature there is Bose-Einstein condensation. Namely, if
$J\geq 0$ is the unique integer defined by the inequalities
\be\label{J}
\veps_J-\veps_0\leq L(U)<\veps_{J+1}-\veps_0\ ,
\ee
for $\beta=\infty$ we have
\be\label{BECint}
\liminf_{N\to\infty}\frac{1}{N}
\sum_{j=0}^J\langle n_j\rangle_{\beta H_N}\geq 1-\frac{L(U)}{\veps_{J+1}-\veps_0}\ .
\ee
(ii) If $U_N$ is a stable integrable pair interaction,
\be\label{pairenergy}
U_N(\r_1,\ldots,\r_N)=\sum_{1\leq i<j\leq N}u_N(\r_i-\r_j)\ ,
\ee
with $\|u_N\|_1=O(1/N)$, then $L(U)<\infty$ and
for any $\beta>0$ there is Bose-Einstein condensation and (\ref{BECint}) holds true.

\end{theorem}

We note that in part (i) $U_N$ can be any self-adjoint operator in $L^2(\R^d)^{\otimes N}$ which is bounded
below and leaves the symmetric subspace invariant.

\vspace{2mm}
\noindent
{\em Proof.}

In the first part of the proof we apply convexity inequalities in a similar manner as they were used in
\cite{LVZ}.

We define a one-parameter family of one-particle Hamiltonians
\be
H^0(\delta)=\sum_{j=0}^J(\veps_j+\delta)P_j+\sum_{j=J+1}^\infty\veps_j P_j
\ee
where $\delta$ is a real parameter and $P_j$ is the orthogonal projection onto $\varphi_j$. Let
$H^0_N(\delta)$ be the corresponding Hamiltonian of $N$ noninteracting particles and
$H_N(\delta)=H^0_N(\delta)+U_N$. For $\delta=0$ we shall keep the original notation.
Because
\be
\sum_{j=0}^J\langle n_j\rangle_{\beta H_N^{(0)}(\delta)}
=-\frac{\partial\ln Z[\beta H_N^{(0)}(\delta)]}{\partial(\beta\delta)}
\ee
and the second derivative is the variance of $\sum_{j=0}^Jn_j$, $\ln Z[\beta H_N^{(0)}(\delta)]$
are convex (decreasing) functions of $\beta\delta$. Therefore, for any $\delta>0$
\be\label{conv1}
\sum_{j=0}^J\langle n_j\rangle_{\beta H_N}\geq \frac{1}{\beta\delta}
\ln\frac{Z[\beta H_N]}{Z[\beta H_N(\delta)]}
\qquad
\sum_{j=0}^J\langle n_j\rangle_{\beta H_N^{0}(\delta)}
\leq \frac{1}{\beta\delta}\ln\frac{Z[\beta H_N^0]}{Z[\beta H_N^{0}(\delta)]}
\ee
Combining (\ref{conv1}) with a double application of the Bogoliubov
convexity inequality \cite{ZB},
\be
\ln\frac{Z[\beta H_N]}{Z[\beta H_N^0]}\geq -\beta\langle U_N\rangle_{\beta H_N^0}\qquad
\ln\frac{Z[\beta H_N^{0}(\delta)]}{Z[\beta H_N(\delta)]}\geq\beta\langle U_N\rangle_{\beta H_N(\delta)}\ ,
\ee
we find
\bea\label{n0int}
\sum_{j=0}^J\langle n_j\rangle_{\beta H_N}
&\geq&
\sum_{j=0}^J\langle n_j\rangle_{\beta H_N^{0}(\delta)}-
\frac{1}{\delta}[\langle U_N\rangle_{\beta H_N^0}-\langle U_N\rangle_{\beta H_N(\delta)}]\nonumber\\
&\geq&
\langle n_0\rangle_{\beta H_N^{0}(\delta)}-
\frac{1}{\delta}[\langle U_N\rangle_{\beta H_N^0}-\inf U_N]\ .
\eea
For $\delta<\veps_{J+1}-\veps_0$, $\varphi_0$ is the unique ground state of $H^0(\delta)$, and thus the
results of Section 3 remain valid to $H^0_N(\delta)$.
At zero temperature the inequality (\ref{n0int}) reads
\be
\lim_{\beta\to\infty}\sum_{j=0}^J\langle n_j\rangle_{\beta H_N}
\geq N-\frac{1}{\delta}[(\Phi_0,U_N\Phi_0)-\inf U_N]\ .
\ee
Dividing by $N$, taking the liminf as $N$ tends to infinity and then letting $\delta$ tend to
$\veps_{J+1}-\veps_0$,
we obtain part (i) of the theorem.

Suppose now that the conditions of part (ii) hold true. Stability means $\inf U_N\geq -BN$ for some
constant $B$. On the other hand,
\bea\label{fiun}
|(\Phi_0,U_N\Phi_0)|={N\choose 2}\left|\int\varphi_0(\x)^2 u_N(\x-\y)\varphi_0(\y)^2\d\x\d\y\right|
\phantom{aaaaaaaaa}\nonumber\\
={N\choose 2}(2\pi)^{d/2}
\left|\int\hat u_N(\q)|\widehat{\varphi_0^2}(\q)|^2\d\q\right|
\leq{N\choose 2}\|u_N\|_1 \|\varphi_0^4\|_1=O(N)\ ,
\eea
therefore
\be
L(U)\leq \frac{1}{2}\|\varphi_0^4\|_1 \limsup\left(N\|u_N\|_1\right)+B<\infty\ .
\ee
Fix $J$ according to (\ref{J}).
Dividing (\ref{n0int}) by $N$, taking the liminf as $N$ tends to infinity and then letting $\delta$ tend to
$\veps_{J+1}-\veps_0$, we arrive at
\be\label{liminf}
\liminf_{N\to\infty}\frac{1}{N}\sum_{j=0}^J\langle n_j\rangle_{\beta H_N}\geq
1-\frac{1}{\veps_{J+1}-\veps_0}\limsup_{N\to\infty}\frac{1}{N}[\langle U_N\rangle_{\beta H_N^0}-\inf U_N]\ .
\ee
Here we used (\ref{Nprime}) to obtain
$\langle n_0\rangle_{\beta H_N^{0}(\delta)}/N=1-\langle N'\rangle_{\beta H_N^{0}(\delta)}/N\to 1$ as
$N$ tends to infinity.
Next we prove that for any $\beta>0$
\be\label{equiv}
\lim_{N\to\infty}\frac{1}{N}[(\Phi_0,U_N\Phi_0)-\langle U_N\rangle_{\beta H_N^0}]=0\ .
\ee
Equations (\ref{liminf}) and (\ref{equiv}) clearly imply (ii).

Let $a(\x)$ and $a(\x)^*$ be the boson field operators, then $U_N$ is the restriction of
\be
U=\frac{1}{2}\int a(\x)^*a(\y)^*u_N(\x-\y)a(\x)a(\y)\d \x \d\y
\ee
to the $N$-particle subspace. Denote $a_i$ and $a_i^*$ the annihilation and creation operators of a
particle in the state $\varphi_i$, respectively. We have
\be
a_i=\int\varphi_i(\x)a(\x)\d\x\quad,\quad a(\x)=\sum_{i=0}^\infty a_i\overline{\varphi_i}(\x)
\ee
and
\be
U=\sum_{i,j,k,l=0}^\infty u_{ijkl}a_i^*a_j^*a_k a_l
\ee
with
\be
u_{ijkl}=\int\varphi_i(\x)\varphi_j(\y)u_N(\x-\y)\overline{\varphi_k}(\x)\overline{\varphi_l}(\y)\d\x\d\y\ .
\ee
Now
\be
\langle\n|U|\n\rangle=\sum_i u_{iiii}{n_i\choose 2}+\sum_{i<j}(u_{ijij}+u_{ijji})n_in_j
\ee
and thus
\bea
\langle U_N\rangle_{\beta H^0_N}&=&\frac{1}{2}\sum_i u_{iiii}\langle n_i(n_i-1)\rangle_{\beta H^0_N}
+\sum_{i<j}(u_{ijij}+u_{ijji})\langle n_in_j\rangle_{\beta H^0_N}\nonumber\\
&=&
\frac{1}{2} u_{0000}\langle n_0(n_0-1)\rangle_{\beta H^0_N}+R_N\nonumber\\
&=&(\Phi_0,U_N\Phi_0)
\left\langle\left(1-\frac{N'}{N}\right)\left(1-\frac{N'}{N-1}\right)\right\rangle_{\beta H^0_N}
+R_N
\eea
where
\be
R_N=\frac{1}{2}\sum_{i>0} u_{iiii}\langle n_i(n_i-1)\rangle_{\beta H^0_N}
+\sum_{0\leq i<j}(u_{ijij}+u_{ijji})\langle n_in_j\rangle_{\beta H^0_N}\ .
\ee
First we estimate the residue $R_N$. Because
\be
\max\{|u_{ijij}|,|u_{ijji}|\}\leq \|\varphi_i\|_\infty \|\varphi_j\|_\infty \|u_N\|_1\ ,
\ee
\bea
|R_N|\leq \|u_N\|_1
\left(\frac{1}{2}\sum_{i>0}\|\varphi_i\|_\infty^2 \langle n_i(n_i-1)\rangle_{\beta H^0_N}\nonumber\right.\\
\left.+2\sum_{0\leq i<j} \|\varphi_i\|_\infty \|\varphi_j\|_\infty\langle n_in_j\rangle_{\beta H^0_N}\right)\ .
\eea
There is some constant $c_1(\beta)$ such that for any $i,j>0$
\bea
\langle n_i(n_i-1)\rangle_{\beta H^0_N}\leq c_1(\beta)e^{-2\beta(\veps_i-\veps_0)}\nonumber\\
\langle n_in_j\rangle_{\beta H^0_N}\leq c_1(\beta)e^{-\beta(\veps_i-\veps_0)}e^{-\beta(\veps_j-\veps_0)}\ .
\eea
These inequalities can be shown by direct estimation.
Also, both expectation values can asymptotically be
computed by using the asymptotic ($N\to\infty$) factorization of the probability measure,
\be
\p0(\{n_j\}_{j>0})\asymp\prod_{j=1}^\infty(1-e^{-\beta(\veps_j-\veps_0)})e^{-\beta(\veps_j-\veps_0)n_j}\ .
\ee
With another suitable constant $c_2(\beta)$ we obtain
\bea
&&\frac{1}{2}\sum_{i>0}\|\varphi_i\|_\infty^2 \langle n_i(n_i-1)\rangle_{\beta H^0_N}
+2\sum_{0<i<j} \|\varphi_i\|_\infty \|\varphi_j\|_\infty\langle n_in_j\rangle_{\beta H^0_N}
\phantom{aaaaaaa}\nonumber\\
&\leq&
c_2(\beta)\left(\sum_{i=1}^\infty \|\varphi_i\|_\infty e^{-\beta(\veps_i-\veps_0)}\right)^2\nonumber\\
&\leq&
c_2(\beta)e^{\beta(\veps_0-V_m)}
\left(\frac{m}{2\pi\hbar^2\beta}\right)^{d/2}\left(\tr e^{-\frac{\beta}{2}(H^0-\veps_0)}-1\right)^2
\equiv \tilde c(\beta)\ .
\eea
In the last inequality we used the bound
\be
\|\varphi_i\|_\infty e^{-\frac{\beta}{2}(\veps_i-\veps_0)}\leq e^{\frac{\beta}{2}(\veps_0-V_m)}
\left(\frac{m}{2\pi\hbar^2\beta}\right)^{d/4}
\ee
obtained in Proposition \ref{2.2}. The remaining term
\bea
2\|\varphi_0\|_\infty\sum_{j=1}^\infty \|\varphi_j\|_\infty\langle n_0n_j\rangle_{\beta H^0_N}
\leq 2N
\|\varphi_0\|_\infty\sum_{j=1}^\infty \|\varphi_j\|_\infty\langle n_j\rangle_{\beta H^0_N}
\leq c(\beta)N
\eea
because for $j>0$, $\langle n_j\rangle_{\beta H^0_N}\sim e^{-\beta(\veps_j-\veps_0)}$.
Hence,
\be\label{rest}
\frac{1}{N}|R_N|\leq \|u_N\|_1\left(c(\beta)+\frac{\tilde c(\beta)}{N}\right)\to 0\quad
\mbox{as $N\to\infty$}\ .
\ee
Therefore,
\bea\label{diff}
\lim_{N\to\infty}\frac{1}{N}[(\Phi_0,U_N\Phi_0)-\langle U_N\rangle_{\beta H_N^0}]
\phantom{aaaaaaaaaaaaaaaaaaaaaaaaaaaaaaaaa}\\
=\lim_{N\to\infty}\frac{1}{N}(\Phi_0,U_N\Phi_0)\left[1-
\left\langle\left(1-\frac{N'}{N}\right)\left(1-\frac{N'}{N-1}\right)\right\rangle_{\beta H^0_N}\right]
=0\nonumber
\eea
because the difference in the square bracket is of order $1/N$, cf. (\ref{nprimk}),
and its prefactor is of order 1, see (\ref{fiun}). This finishes the proof of the theorem.

\vspace{2mm}
Notice that in the proof of (\ref{rest}) and (\ref{diff}) we have used only $\|u_N\|_1=o(1)$.
The condition of integrability of $u_N$ could be relaxed.
For example, if $u$ is a bounded function (or $u$ is integrable and bounded below), the theorem
holds for $u_N=(1/N)u$, which is a mean-field interaction.
More interesting examples are provided by scaled interactions.

\begin{coro}
Let $u:\R^d\to\R$ be an integrable nonnegative function. Suppose we are given 
two positive sequences $\alpha_N$ and $b_N$ satisfying the condition
\be\label{cond3}
S\equiv\sup_{N}\ b_N\alpha_N^{-d} N<\infty\ .
\ee
Then for any $\beta>0$
there is Bose-Einstein condensation for the interaction
\be\label{uN}
u_N(\x)=b_Nu(\alpha_N\x)\ .
\ee
\end{coro}

\noindent
{\em Proof.}
$U_N$ is an integrable stable pair interaction ($\inf U_N=0$) and
\be\label{uNbound}
\|u_N\|_1=b_N\alpha_N^{-d}\|u\|_1\leq\frac{S\|u\|_1}{N}\ .
\ee
Thus, the conditions of part (ii) of the theorem are verified.

\vspace{2mm}
\noindent
{\em Remarks.}\\
{\bf 1.}
If $\alpha_N$ is constant, we obtain the mean-field interaction.
If $\alpha_N$ is strictly monotonous, it can be
inverted and, hence, $b_N$ may depend on $N$ only via $\alpha_N$. 
For example,
$\alpha_N=N^{\eta}$ and $b_N=\alpha_N^{d-1/\eta}$ satisfy (\ref{cond3}).
\\
{\bf 2.} If the scattering length of $u$ is $a$ and $b_N=\alpha_N^2$ then
the scattering length of $u_N$ is $a/\alpha_N$. To see this,
we recall (cf. \cite{LY}) the definition of the scattering length:

Let $V$ be a spherical finite-range potential such that $-\h2m\Delta+V$ has no negative or
zero energy
bound state. Then the Schr\"odinger equation written for zero energy,
\be\label{Sch}
-\h2m\Delta\phi(\x)+V(\x)\phi(\x)=0
\ee
has a (up to constant multipliers) unique spherical sign-keeping solution, $\phi_0$. If $r=|\x|>R_0$,
the range of the potential, this solution reads
\bea\label{defscatt}
\phi_0(\x)=\left\{\begin{array}{ll}
       1-(a/r)^{d-2}       &    \mbox{if $d\neq 2$}\\
       \ln(r/a)  &    \mbox{if $d=2$}\\
              \end{array}
       \right.
\eea
with some $a\leq R_0$.
We call $a$ the scattering length
of $V$ and $\phi_0$ the defining solution. To obtain the scattering length of a pair interaction $u$ one
has to solve (\ref{Sch}) with $V=u/2$, the 1/2 accounting for the reduced mass.
For a nonnegative integrable infinite range potential (pair interaction)
a finite scattering length still can be defined by truncating the potential at a finite $R_0$
and taking the (finite) limit of $a(R_0)$ as $R_0\to\infty$, see Appendix A of \cite{LY}.

Suppose now that the scattering length of $u$ is $a$. What is the scattering length of
$u_N$, given by (\ref{uN})? This is not always easy to tell because the defining solution for $u_N$ is
generally in no simple relation with that one for $u$. However, from equations (\ref{Sch})
and (\ref{defscatt}) it is easily seen that the defining
solutions of $u$ and
\be\label{ua}
u_\alpha(\x)=\alpha^2u(\alpha\x)
\ee
are related by scaling, $\phi_0[u_\alpha](\x)=\phi_0[u](\alpha\x)$, and therefore the
scattering length of $u_\alpha$ is $a/\alpha$.
\\
{\bf 3.}
If $\alpha_N$ tends to infinity, the scattering length of $u_N$ tends
to zero, and the operator $-\h2m\Delta+u_N/2$ converges in norm resolvent sense to
the one-particle kinetic energy operator. For this to happen, in two and three dimensions
$u_N\geq 0$ is essential. Indeed,
in two and three dimensions with $\alpha_N$ diverging and $b_N$ chosen so that (\ref{cond3})
is respected one could
define point interactions, that is, self-adjoint
extensions of the symmetric operator $-\h2m\Delta|_{C_0^\infty(\R^d-\{0\})}$ with a nonvanishing
scattering length \cite{AGHH}. However, it turns out that for $u_N\geq 0$ the only extension
is the kinetic energy operator (cf. Theorems 1.2.5 and 5.5 of \cite{AGHH}). 
The result of Theorem 2 and its
corollary can be nontrivial because the scattering length vanishes in conjunction with a diverging
particle number.
\\
{\bf 4.}
In three dimensions the Gross-Pitaevskii scaling limit is obtained by fixing $Na_N$, where $a_N$ is the
scattering length of the pair interaction, while $N\to\infty$. To show BEC, we choose
$b_N=\alpha_N^2$ and $\alpha_N\propto N$, so that $u_N=u_{\alpha_N}$ with scattering length $a_N=a_1/N$. Observe that
$\|u_{\alpha_N}\|_1=N^{-1}\|u\|_1$ for GP scaling in three dimensions.
\\
{\bf 5.}
Let $H_{N}[V,u]$ be an $N$-particle Hamiltonian with an external potential $V$
and pair interaction $u$. Then
\begin{equation}\label{1}
\beta
H_{N}[V,\alpha_N^2u(\alpha_N\,\cdot)]
\cong\beta\alpha_N^2H_{N}[\alpha_N^{-2}V(\alpha_N^{-1}\,\cdot),u]\ .
\end{equation}
If $\alpha_N$ tends to infinity, the scaled temperature $(\beta\alpha_N^2)^{-1}$ goes to zero
and the trap opens. Lieb and Seiringer \cite{LS} obtained
results on the limit of the sequence of ground states of (\ref{1}). Theorem 2
refers to the limit of the thermal equilibrium states generated by (\ref{1}).
It is not obvious that the two limits define the same state for $N=\infty$.
In Theorem 2 there is a first hint that this may hold true: By proving equation (\ref{equiv}),
we obtain the same lower bound (\ref{BECint}) on the density of the
condensate at positive temperatures as at zero temperature.
\\
{\bf 6.} In two dimensions the scattering length of $u_N$ is always smaller than
$a/\alpha_N$, the scattering length of $u_{\alpha_N}$,
cf. (\ref{ua}). In general,
\be\label{uNa}
u_N(\x)=b_N\alpha_N^{-2}u_{\alpha_N}(\x)\leq S N^{-1}\alpha_N^{d-2}u_{\alpha_N}(\x)\ .
\ee
In particular, in two dimensions $u_N\leq (S/N)u_{\alpha_N}$.
Because for $u\geq 0$ the scattering length of $\lambda u$
increases with $\lambda>0$, the scattering length of any admissible $u_N$ is smaller than that of
$u_{\alpha_N}$.
We note that in two dimensions $\|u_{\alpha}\|_1=\|u\|_1$, independently of $\alpha$.
\\
{\bf 7.}
The sequence $\alpha_N$ may also decrease with $N$, provided that $b_N$ decreases sufficiently
rapidly, see e.g. Remark 1 for $\eta<0$. A curious example in one dimension is $\alpha_N=N^{-1}$ and 
$b_N=N^{-2}$.
Thus, the scattering length increases proportional to $N$, instead of going to zero. According to
equation (\ref{1}), this case corresponds to closing
the trap and sending the temperature to infinity -- just the opposite of the
Gross-Pitaevskii limit in three dimensions.
\\
{\bf 8.}
Theorem 2 is
valid for a gas confined in a box with periodic, Neumann or Dirichlet
boundary conditions. The geometric confinement on a $d$-torus is interesting
because the eigenstates of the one-particle Hamiltonian are eigenstates of the
one-particle reduced density matrix as well, see Section 4.1. Now the inequality
(\ref{BECint}) implies that at least $\varphi_0$ is macroscopically occupied and suggests
that $\langle n_j\rangle$ for some small positive $j$ can also be of order $N$.
This would mean a kind of generalized Bose-Einstein condensation, in contrast to
the 100\% condensation into a single state, obtained for locally bounded trap
potentials \cite{LS}.
\\
{\bf 9.}
For bosons in a locally bounded potential trap scaling of a nonnegative interaction is
unavoidable in order that condensation takes place into a fixed localized state $\varphi$.
Particles in $\varphi$ are
confined in a bounded box with a probability arbitrarily close to 1.
An unscaled nonnegative interaction would push the particles outside this box and, hence, out of $\varphi$.
In effect, with increasing $N$ the system could diminish its interaction energy at the expense
of the potential energy, by letting the particles "climb" a little bit higher up in the
potential well.


\vspace{2mm}
\noindent
{\bf Acknowledgment}\\
The manuscript underwent a substantial revision in response to some pertinent
remarks of Philippe Martin and Robert Seiringer. I thank Barry Simon for a correspondence
on ultracontractivity.
This work was supported by the Hungarian Scientific Research Fund (OTKA)
through grants T 030543 and T 042914.

\end{document}